\documentclass[12pt,draftcls, onecolumn,journal]{IEEEtran}
\usepackage{times}
\usepackage[final]{graphicx}
\usepackage[reqno]{amsmath}
\usepackage{amsfonts}
\usepackage{times,amsmath,epsfig}
\usepackage{latexsym,amssymb}
\usepackage{cite}

\def\myQED{\mbox{\rule[0pt]{1.5ex}{1.5ex}}}

\newtheorem{thm}{Theorem}

\newtheorem{lem}[thm]{Lemma}

\newtheorem{define}[thm]{Definition}

\newcommand{\no}{\nonumber}

\begin{document}

\title{The Wiretap Channel with Feedback: Encryption over the Channel}

\author{
Lifeng Lai, Hesham El Gamal and H. Vincent Poor
\thanks{Lifeng Lai (llai@princeton.edu) was with
the Department of Electrical and Computer Engineering at the Ohio
State University, he is now is with the Department of Electrical
Engineering at Princeton University. Hesham El Gamal
(helgamal@ece.osu.edu) is with the Department of Electrical and
Computer Engineering at the Ohio State University. H. Vincent Poor
(poor@princeton.edu) is with the Department of Electrical
Engineering at Princeton University. This research was supported
by the National Science Foundation under Grants ANI-03-38807 and
CNS-06-25637.}}
\date{}
\maketitle 



\begin{abstract}
In this work, the critical role of noisy feedback in enhancing the
secrecy capacity of the wiretap channel is established. Unlike
previous works, where a noiseless public discussion channel is used
for feedback, the feed-forward and feedback signals share the same
noisy channel in the present model. Quite interestingly, this noisy
feedback model is shown to be more advantageous in the current
setting. More specifically, the discrete memoryless modulo-additive
channel with a full-duplex destination node is considered first, and
it is shown that the judicious use of feedback increases the perfect
secrecy capacity to the capacity of the source-destination channel
in the absence of the wiretapper. In the achievability scheme, the
feedback signal corresponds to a private key, known only to the
destination. In the half-duplex scheme, a novel feedback technique
that always achieves a positive perfect secrecy rate (even when the
source-wiretapper channel is less noisy than the source-destination
channel) is proposed. These results hinge on the modulo-additive
property of the channel, which is exploited by the destination to
perform encryption over the channel without revealing its key to the
source. Finally, this scheme is extended to the continuous real
valued modulo-$\Lambda$ channel where it is shown that the perfect
secrecy capacity with feedback is also equal to the capacity in the
absence of the wiretapper.

\end{abstract}

\section{Introduction} \label{sec:intro}

The study of secure communication from an information theoretic
perspective was pioneered by Shannon~\cite{Shannon:BSTJ:49}. In
Shannon's model, both the sender and the destination possess a
common secret key $K$, which is unknown to the wiretapper, and use
this key to encrypt and decrypt the message $M$. Shannon considered
a scenario where both the legitimate receiver and the wiretapper
have direct access to the transmitted signal and introduced the
perfect secrecy condition $I(M;Z)=0$, implying that the signal $Z$
received by the wiretapper does not provide any additional
information about the source message $M$. Under this model, he
proved the pessimistic result that the achievability of perfect
secrecy requires the entropy of the shared private key $K$ to be at
least equal to the entropy of the message itself (i.e., $H(K) \ge
H(M)$ for perfect secrecy). Clearly, the distribution of the secret
key under this model is challenging.

In a pioneering work~\cite{Wyner:BSTJ:75}, Wyner introduced the
wiretap channel and established the possibility of creating an
almost perfectly secure source-destination link without relying on
private (secret) keys. In the wiretap channel, both the wiretapper
and destination observe the source encoded message through noisy
channels. Similar to Shannon's model, the wiretapper is assumed to
have unlimited computational resources. Wyner showed that when the
source-wiretapper channel is a degraded version of the
source-destination channel, the source can send perfectly
secure\footnote{Wyner's notion of per symbol equivocation is weaker
than Shannon's notion of perfect secrecy~\cite{Maurer:LNCS:00}.}
messages to the destination at a non-zero rate. The main idea is to
hide the information stream in the additional noise impairing the
wiretapper by using a stochastic encoder which maps each message to
many codewords according to an appropriate probability distribution.
This way, one induces maximal equivocation at the wiretapper. By
ensuring that the equivocation rate is arbitrarily close to the
message rate, one achieves perfect secrecy in the sense that the
wiretapper is now limited to learn {\em almost nothing} about the
source-destination messages from its observations. Follow-up work by
Leung-Yan-Cheong and Hellman has characterized the secrecy capacity
of the additive white Gaussian noise (AWGN) wiretap
channel~\cite{Leung:TIT:78}. In a landmark paper, Csisz$\acute{a}$r
and K\"{o}rner generalized Wyner's approach by considering the
transmission of confidential messages over broadcast
channels~\cite{Csiszar:TIT:78}. This work characterized the perfect
secrecy capacity of Discrete Memoryless Channels (DMC)s, and showed
that the perfect secrecy capacity is positive unless the
source-wiretapper channel is {\em less noisy} than the
source-destination channel (referred to as the main channel in the
sequel)\footnote{The source-wiretapper channel is said to be less
noisy than the main channel, if for every $V\to X\to YZ$,
$I(V;Z)\geq I(V;Y)$, where $X$ is the signal transmitted by the
source, and where $Y$ and $Z$ are the received signal at the
receiver and the wiretapper respectively.}.

Positive secrecy capacity is not always possible to achieve in
practice. In an attempt to transmit messages securely in these
unfavorable scenarios, \cite{Maurer:TIT:93} and
\cite{Ahlswede:TIT:93} considered the wiretap channel with
\textbf{noiseless} feedback\footnote{The authors also considered a
more general secret sharing problem. }. They showed that one may
leverage the feedback to achieve a positive perfect secrecy rate,
even when the feed-forward perfect secrecy capacity is zero. In this
model, there exists a separate noiseless public channel, through
which the transmitter and receiver can exchange information. The
wiretapper is assumed to obtain a perfect copy of the messages
transmitted over this public channel. Upper and lower bounds were
derived for the perfect secrecy capacity with noiseless feedback
in~\cite{Maurer:TIT:93,Ahlswede:TIT:93}. In several cases, as
discussed in detail in the sequel, these bounds coincide. But, in
general, the perfect secrecy capacity with noiseless feedback
remains unknown. Along the same line, \cite{Csiszar:TIT:00}
established the critical role of a trusted/untrusted helper in
enhancing the secret key capacity of public discussion algorithms.
The multi-terminal generalization of the basic set-up of
\cite{Maurer:TIT:93,Ahlswede:TIT:93} was studied in
\cite{Csiszar:TIT:04}. Finally,
in~\cite{Maurer:TIT:03,Maurer:TIT:031,Maurer:TIT:032}, the public
discussion paradigm was extended to handle the existence of active
adversaries.

Our work represents a marked departure from the public discussion
paradigm. In our model, we do not assume the existence of a
separate noiseless feedback channel. Instead, the feedback signal
from the destination, which is allowed to depend on the signal
received so far, is transmitted over the same noisy channel used
by the source. Based on the noisy feedback signal, the source can
then causally adapt its transmission scheme, hoping to increase
the perfect secrecy rate. The wiretapper receives a mixture of the
signal from the source and the feedback signal from the
destination. Quite interestingly, we show that in the
modulo-additive DMC with a full-duplex destination, the perfect
secrecy capacity with noisy feedback equals the capacity of the
main channel in the absence of the wiretapper. Furthermore, the
capacity is achieved with a simple scheme where the source ignores
the feedback signal and the destination feeds back randomly
generated symbols from a certain alphabet set. This feedback
signal plays the role of a private key, known only by the
destination, and encryption is performed by the modulo-additive
channel. The more challenging scenario with a half-duplex
destination, which cannot transmit and receive simultaneously, is
considered next. Here, the active transmission periods by the
destination will introduce erasures in the feed-forward
source-destination channel. In this setting, we propose a novel
feedback scheme that achieves a positive perfect secrecy rate for
any non-trivial channel distribution. The feedback signal in our
approach acts as a private {\em destination only} key which
strikes the optimal tradeoff between introducing erasures at the
destination and errors at the wiretapper. Finally, the proposed
scheme is extended to the continuous modulo-$\Lambda$ lattice
channel where it is shown to achieve the capacity of the main
channel. Overall, our work proposes a novel approach for
encryption where 1) the feedback signal is used as a private key
known only to the destination and 2) the encryption is performed
by exploiting the modulo-additive property of the channel. This
encryption approach is shown to be significantly superior to the
classical public discussion paradigm.

Recently, there has been a resurgent interest in studying secure
communications from information theoretic perspective under
various scenarios. The point-to-point fading eavesdropper channel
was considered
in~\cite{Gopala:TIT:06,Liang:TIT:061,Bloch:TIT:06,Bloch:TIT:061,Li:ITA:07,Parada:ISIT:05}
under different assumptions on the delay constraints and the
available transmitter Channel State Information (CSI).
In~\cite{Tekin:TIT:06,Tekin:TIT:07,Liang:TIT:06,Liu:ISIT:06}, the
information theoretic limits of secure communications over
multiple access channels were explored. The relay channel with
confidential messages, where the relay acts both as a wiretapper
and a helper, was studied in~\cite{Oohama:ITW:01,Oohama:TIT:06}.
In~\cite{Liu:TIT:07}, the interference channel with confidential
messages was studied. In~\cite{Lai:TIT:061}, the four terminal
relay-eavesdropper channel was introduced and analyzed. The
wiretap channel with side information was studied
in~\cite{Mitrpant:TIT:06}.

The rest of the paper is organized as follows. In
Section~\ref{sec:model}, we introduce the system model and our
notation. Section~\ref{sec:full} describes and analyzes the
proposed feedback scheme which achieves the capacity of the full
duplex modulo-additive DMC. Taking the Binary Symmetric Channel
(BSC) as an example, we then compare the performance of the
proposed scheme with the public discussion approach. The
half-duplex scenario is studied in Section~\ref{sec:half}. In
Section~\ref{sec:exten}, we extend our results to the
modulo-$\Lambda$ lattice channel. Finally, Section~\ref{sec:con}
offers some concluding remarks and outlines possible venues for
future research.

\section{The Modulo-Additive Discrete Memoryless Channel}\label{sec:model}
Throughout the sequel, the upper-case letter $X$ will denote a
random variable, a lower-case letter $x$ will denote a realization
of the random variable, a calligraphic letter $\mathcal{X}$ will
denote a finite alphabet set and a boldface letter $\mathbf{x}$
will denote a vector. Furthermore, we let $[x]^+=\max\{0,x\}$.
Without feedback, our modulo-additive discrete memoryless wiretap
channel is described by the following relations at time $i$
\begin{eqnarray}
y(i)=x(i)+n_1(i),\no\\
z(i)=x(i)+n_2(i),
\end{eqnarray}
where $y(i)$ is the received symbol at the destination, $z(i)$ is
the received symbol at the wiretapper, $x(i)$ is the channel
input, $n_1(i)$ and $n_2(i)$ are the noise samples at the
destination and wiretapper, respectively. Here $N_1$ and $N_2$ are
allowed to be correlated, while each process is assumed to be
individually drawn from an identically and independently
distributed source. Also we have
$X\in\mathcal{X}=\{0,1,\cdots,|\mathcal{X}|-1\},Y,N_1\in\mathcal{Y}=\{0,1,\cdots,|\mathcal{Y}|-1\}$
and $Z,N_2\in\mathcal{Z}=\{0,1,\cdots,|\mathcal{Z}|-1\}$ with
finite alphabet sizes $|\mathcal{X}|,|\mathcal{Y}|,|\mathcal{Z}|$
respectively. Here `$+$' is understood to be modulo addition with
respect to the corresponding alphabet size, i.e.,
$y(i)=[x(i)+n_1(i)]\mod |\mathcal{Y}|$ and $z(i)=[x(i)+n_2(i)]\mod
|\mathcal{Z}|$ with addition in the real field.

In this paper, we focus on the wiretap channel with noisy
feedback. More specifically, at time $i$ the destination sends the
causal feedback signal $X_{1}(i)$ over the same noisy channel used
for feed-forward transmission, \emph{i.e.}, we do not assume the
existence of a separate noiseless feedback channel. The causal
feedback signal is allowed to depend on the received signal so far
$Y^{i-1}$, \emph{i.e.}, $X_{1}(i)=\Psi (Y^{i-1})$, where $\Psi$
can be any (possibly stochastic) function. In general, we allow
the destination to choose the alphabet of the feedback signal
$\mathcal{X}_1$ and the corresponding size $|\mathcal{X}_1|$. With
this {\em noisy} feedback from the destination, the received
signal at the source, wiretapper and destination are
\begin{eqnarray}
y_{0}(i)=x(i)+x_{1}(i)+n_0(i),\no\\
y(i)=x(i)+x_{1}(i)+n_{1}(i),\no
\end{eqnarray}
and
$$
z(i)=x(i)+x_{1}(i)+n_2(i),\no $$ respectively. Here
$Y_{0}\in\mathcal{Y}_0=\{0,1,\cdots,|\mathcal{Y}_0|-1\}$ is the
received noisy feedback signal at the source and $N_0$ is the
feedback noise, which may be correlated with $N_1$ and $N_2$. We
denote the alphabet size of $N_0$ and $Y_0$ by $|\mathcal{Y}_0|$.
Again, all `$+$' operation should be understood to be modulo
addition with corresponding alphabet size.

Now, the source wishes to send the message $W\in
\mathcal{W}=\{1,\cdots,M\}$ to the destination using a $(M,n)$ code
consisting of: 1) a casual stochastic encoder $f$ at the source that
maps the message $w$ and the received noisy feedback signal
$y_0^{i-1}$ to a codeword $\mathbf{x}\in \mathcal{X}^{n}$ with
\begin{eqnarray}
x(i)=f(i,w,y_0^{i-1}),
\end{eqnarray}
2) a stochastic feedback encoder $\Psi$ at the destination that
maps the received signal into $X_1(i)$ with $x_1(i)=\Psi(y^{i-1})$
and 3) a decoding function at the destination $d$:
$\mathcal{Y}^{n}\rightarrow \mathcal{W}$. The average error
probability of the $(M,n)$ code is
\begin{eqnarray}
P_{e}^{n}=\sum\limits_{w\in\mathcal{W}}\frac{1}{M}\text{Pr}\{d(\mathbf{y})\neq
w|w\text{ was sent}\}.
\end{eqnarray}
The equivocation rate at the wiretapper is defined as
\begin{eqnarray}
R_{e}=\frac{1}{n}H(W|\mathbf{Z}).
\end{eqnarray}

We are interested in perfectly secure transmission rates defined
as follows.

\begin{define}
A secrecy rate $R^{f}$ is said to be achievable over the wiretap
channel with noisy feedback if for any $\epsilon>0$, there exists
a sequence of codes $(M,n)$ such that for any $n\geq n(\epsilon)$,
we have
\begin{eqnarray}
R^{f}&=&\frac{1}{n}\log_2M,\label{eq:rate}\\
P_{e}^{n}&\leq& \epsilon,\label{eq:error}\\
\frac{1}{n}H(W|\mathbf{Z})&\geq& R^{f}-\epsilon.\label{eq:eqvo}
\end{eqnarray}
\end{define}

\begin{define}
The secrecy capacity with noisy feedback $C_s^{f}$ is the maximum
rate at which messages can be sent to the destination with perfect
secrecy; \emph{i.e.}
\begin{eqnarray}
C_s^{f}=\sup\limits_{f,\Psi}\{R^{f}:R^{f}\;\; \text{is
achievable}\}.
\end{eqnarray}
\end{define}

Note that in our model, the wiretapper is assumed to have unlimited
computation resources and to know the coding scheme of the source
and the feedback function $\Psi$ used by the destination. We believe
that our feedback model captures realistic scenarios where the
terminals exchange information over noisy channels.

\section{The Wiretap Channel with Full-Duplex Feedback}\label{sec:full}

\subsection{Known Results}

The secrecy capacity of the wiretap DMC without feedback $C_s$ was
characterized in~\cite{Csiszar:TIT:78}. Specializing to our
modulo-additive channel, one obtains
\begin{eqnarray}\label{eq:seccap}
C_{s}=\max\limits_{V\rightarrow X\rightarrow YZ}
[I(V;Y)-I(V;Z)]^+.
\end{eqnarray}

The wiretap DMC with public discussion was introduced and analyzed
in~\cite{Maurer:TIT:93,Ahlswede:TIT:93}. More specifically, these
papers considered a more general model in which all the nodes
observe correlated variables\footnote{ The wiretap channel model
is a particular mechanism for the nodes to observe the correlated
variables, and corresponds to the ``channel type model'' studied
in~\cite{Ahlswede:TIT:93}.}, and there exists an extra noiseless
public channel with infinite capacity, through which both the
source and the destination can send information. Combining the
correlated variables and the publicly discussed messages, the
source and the destination generate a key about which the wiretap
only has negligible information. Please refer
to~\cite{Ahlswede:TIT:93} for rigorous definitions of these
notions. Since the public discussion channel is noiseless, the
wiretapper is assumed to observe a noiseless version of the
information transmitted over it. It is worth noting that some of
the schemes proposed in~\cite{Maurer:TIT:93,Ahlswede:TIT:93}
manage only to generate an identical secret key at both the source
and destination. The source may then need to encrypt its message
using the one-time pad scheme which reduces the effective
source-destination information rate. Thus, the \emph{effective}
secrecy rate that could be used to transmit information from the
source to the destination \emph{may} be less than the results
reported in~\cite{Maurer:TIT:93,Ahlswede:TIT:93}. Nevertheless, we
use these results for comparison purposes (which is generous to
the public discussion paradigm). The following theorem gives upper
and lower bounds on the secret key capacity of the public
discussion paradigm $C_s^{p}$.

\begin{thm}[\cite{Maurer:TIT:93,Ahlswede:TIT:93}] The secret key capacity of the
public discussion approach satisfies the following conditions:
\begin{eqnarray}
\max
\{\max\limits_{P_X}[I(X;Y)-I(X;Z)],\max\limits_{P_X}[I(X;Y)-I(Y;Z)]\}\leq
C_s^{p}\leq \min \{\max\limits_{P_X}I(X;Y),
\max\limits_{P_X}I(X;Y|Z)\}.\no
\end{eqnarray}
\end{thm}
\begin{proof}
Please refer to~\cite{Maurer:TIT:93,Ahlswede:TIT:93}.
\end{proof}

These bounds are known to be tight in the following
cases~\cite{Maurer:TIT:93,Ahlswede:TIT:93}.
\begin{enumerate}

\item $P_{YZ|X}=P_{Y|X}P_{Z|X}$, \emph{i.e.}, the main channel and
the source-wiretapper channel are independent; in this case
\begin{eqnarray}C_s^{p}=\max\limits_{P_X}\{I(X;Y)-I(Y;Z)\}.\label{eq:cwithfin}\end{eqnarray}

\item $P_{XZ|Y}=P_{X|Y}P_{Z|Y}$, \emph{i.e.}, $X\rightarrow
Y\rightarrow Z$ forms a Markov chain, and hence the
source-wiretapper channel is a degraded version of the main
channel. In this case
\begin{eqnarray}
C_{s}^{p}=\max\limits_{P_X}\{ I(X;Y)-I(X;Z)\}\label{eq:cwithfdeg}.
\end{eqnarray}

This is also the secrecy capacity of the degraded wiretap channel
without feedback. Hence public discussion does not increase the
secrecy capacity for the degraded wiretap channel.

\item $P_{XY|Z}=P_{X|Z}P_{Y|Z}$, \emph{i.e.}, $X\rightarrow
Z\rightarrow Y$, so that the main channel is a degraded version of
the wiretap channel. In this case
\begin{eqnarray}\label{eq:maindegrad}
C_s^{p}=0.
\end{eqnarray}

Again, public discussion does not help in this scenario.
\end{enumerate}

\subsection{The Main Result}

Before presenting the main theorem, we present the crypto lemma
which will be intensively used later.

\begin{lem}[Crypto Lemma~\cite{Forney:ALL:03}]\label{lem:crypto}
Let $G$ be a compact abelian group with group operation `+', and
let $Y = X + X_1$, where $X$ and $X_1$ are random variables over
$G$ and $X_1$ is independent of $X$ and uniform over $G$. Then $Y$
is independent of $X$ and uniform over $G$.
\end{lem}
\begin{proof}
Please refer to~\cite{Forney:ALL:03}.
\end{proof}

The following theorem characterizes the secrecy capacity of the
wiretap channel with noisy feedback. Moreover, achievability is
established through a novel encryption scheme that exploits the
modulo-additive structure of the channel and uses a private key
known only to the destination.

\begin{thm}\label{thm:dmc}
The secrecy capacity of the discrete memoryless modulo-additive
wiretap channel with noisy feedback is
\begin{eqnarray}C_s^{f}=C,\end{eqnarray}
where $C$ is the capacity of the main channel in the absence of the
wiretapper.
\end{thm}
\begin{proof}

1. Converse.

Let
\begin{eqnarray}
\mathcal{R}^{f}=\{R^{f}: \text{there exists a coding scheme that
satisfies \eqref{eq:rate}-\eqref{eq:eqvo} for $R^{f}$}\}.
\end{eqnarray}

Also, let
\begin{eqnarray}
\mathcal{R}=\{R: \text{there exists a coding scheme that satisfies
\eqref{eq:rate}-\eqref{eq:error} for $R$}\}.
\end{eqnarray}

Obviously $\mathcal{R}^{f}\subseteq\mathcal{R}$, since we are
dropping off the equivocation condition~\eqref{eq:eqvo},
\emph{i.e.}, we are ignoring the wiretapper. Hence we have
$C_{s}^{f}=\sup\mathcal{R}^{f}\leq \sup \mathcal{R}$. It is clear
that $\mathcal{R}$ is the set of reliable transmission rate of an
ordinary DMC channel with feedback. It is well known that feedback
does not increase the capacity of discrete memoryless channels,
hence we have
\begin{eqnarray}
C_s^{f}=\sup\mathcal{R}^{f}\leq \sup \mathcal{R}= C.
\end{eqnarray}

2. Achievability.

For any given input probability mass function $p(x)$, we use the
following scheme.
\begin{enumerate}
\item Coding at the source.

The source ignores the feedback signal and uses a channel coding
scheme for the ordinary channel without wiretapper. More
specifically, the source generates $M=2^{R^{f}}$ length-$n$
codewords $\mathbf{x}$ with probability
$$p(\mathbf{x})=\prod\limits_{i=1}^{n}p(x(i)).$$
When the source needs to send message $w\in\mathcal{W}$, it sends
the corresponding codeword $\mathbf{x}(w)$.

\item Feedback at the destination.

The destination sets $\mathcal{X}_1=\mathcal{Z}$, and at any time
$i$ sets $x_{1}(i)=a, a\in\{0,\cdots,|\mathcal{Z}|-1\}$ with
probability $1/|\mathcal{Z}|$. Hence $\mathbf{x}_1$ is uniformly
distributed over $\mathcal{Z}^n$.

\item Decoding at the destination.

After receiving $\mathbf{y}$, the destination sets
$\hat{\mathbf{y}}=\mathbf{y}-\mathbf{x}_1$, here `$-$' is
understood to be a component-wise modulo $|\mathcal{Y}|$
operation. It is easy to see that
$\hat{\mathbf{y}}=\mathbf{x}+\mathbf{n}_1$. The destination then
claims that $\hat{w}$ was sent, if
$(\hat{\mathbf{y}},\mathbf{x}(\hat{w}))$ are jointly typical. For
any given $\epsilon>0$, the probability that $\hat{w}\neq w$ goes
to zero, if $R^{f}= I(X;\hat{Y})-\epsilon=I(X;Y|X_1)-\epsilon$ and
$n$ is large enough. The channel $X\rightarrow\hat{Y}$ is
equivalent to the main channel without feedback. Hence as long as
$R^{f}<C$, there exists a code with sufficient code-length such
that $P^{n}_e\leq \epsilon$ for any $\epsilon>0$.

\item Equivocation at the wiretapper.

The wiretapper will receive
\begin{eqnarray}\mathbf{z}=\mathbf{x}+\mathbf{x}_1+\mathbf{n}_2,\end{eqnarray}
and $\mathbf{x}_1$ is uniformly distributed over $\mathcal{Z}^n$
and is independent with $\mathbf{x}$. Based on the crypto lemma,
for any given $\mathbf{x}$, $\mathbf{x}+\mathbf{X}_1$ is uniformly
distributed over $\mathcal{Z}^n$, and hence $\mathbf{z}$ is
uniformly distributed over $\mathcal{Z}^n$ for any transmitted
codeword $\mathbf{x}$ and noise realization $\mathbf{n}_2$.
Moreover $\mathbf{Z}$ is independent with $\mathbf{X}$, thus
\begin{eqnarray}
I(\mathbf{X};\mathbf{Z})=0.
\end{eqnarray}
Hence we have $I(W;\mathbf{Z})\leq I(\mathbf{X};\mathbf{Z})=0$,
thus
\begin{eqnarray}
\frac{1}{n}H(W|\mathbf{Z})=\frac{H(W)-I(W;\mathbf{Z})}{n}=R^{f},
\end{eqnarray}
and we achieve perfect secrecy.
\end{enumerate}
This completes the proof.
\end{proof}

The following observations are now in order.
\begin{enumerate}

\item Our scheme achieves $I(W;\mathbf{Z})=0$. This implies
perfect secrecy in the strong sense of
Shannon~\cite{Shannon:BSTJ:49} as opposed to Wyner's notion of
perfect secrecy~\cite{Wyner:BSTJ:75}, which has been pointed out
to be insufficient for certain encryption
applications~\cite{Maurer:LNCS:00}.

\item The enabling observation behind our achievability scheme is
that, by judiciously exploiting the modulo-additive structure of
the channel, one can render the channel output at the wiretapper
independent from the codeword transmitted by the source. Here, the
feedback signal $\mathbf{x}_1$ serves as a private key and the
encryption operation is carried out by the channel. Instead of
requiring both the source and destination to know a common
encryption key, we show that only the destination needs to know
the encryption key, hence eliminating the burden of secret key
distribution.

\item Remarkably, the secrecy capacity with {\em noisy} feedback
is shown to be larger than the secret key capacity of public
discussion schemes. This point will be further illustrated by the
binary symmetric channel example discussed next. This presents a
marked departure from the conventional {\em wisdom}, inspired by
the data processing inequality, which suggests the superiority of
noiseless feedback. This result is due to the fact that the
noiseless feedback signal is also available to the wiretapper,
while in the proposed noisy feedback scheme neither the source nor
the wiretapper knows the feedback signal perfectly. In fact, the
source in our scheme ignores the feedback signal, which is used
primarily to {\em confuse} the wiretapper.



\item Our result shows that complicated feedback functions $\Psi$
are not needed to achieve optimal performance in this setting (i.e.,
a random number generator suffices). Also, the alphabet size of the
feedback signal can be set equal to the alphabet size of the
wiretapper channel and the coding scheme used by the source is the
same as the one used in the absence of the wiretapper.

\end{enumerate}
\subsection{The Binary Symmetric Channel Example}
\begin{figure}[thb]
\centering
\includegraphics[width=0.2\textwidth]{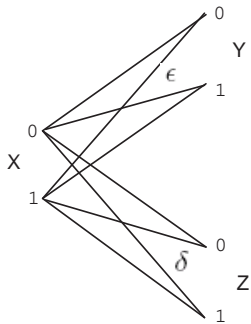}
\caption{The Binary Symmetric Wiretap Channel.} \label{fig:bsc}
\end{figure}
To illustrate the idea of encryption over the channel, we consider
in some details the wiretap BSC shown in Figure~\ref{fig:bsc},
where $\mathcal{X}=\mathcal{Y}=\mathcal{Z}=\{0,1\}$,
$\text{Pr}\{n_1=1\}=\epsilon$ and $\text{Pr}\{n_2=1\}=\delta$. The
secrecy capacity of this channel without feedback is known to
be~\cite{Maurer:TIT:93}
$$C_s=[H(\delta)-H(\epsilon)]^+,$$ with $H(x)=-x\log x-(1-x)\log
(1-x)$. We differentiate between the following special cases.

\begin{enumerate}

\item $\epsilon=\delta=0$.

In this case, both the main channel and wiretap channel are
noiseless, hence $$C_s=0.$$ Also we have $$C_s^p=0,$$ since the
wiretapper sees exactly the same as what the destination sees.
Specializing our scheme to this BSC channel, at time $i$, the
destination randomly chooses $x_1(i)=1$ with probability 1/2 and
sends $x_1(i)$ over the channel. This creates a virtual BSC at the
wiretapper with $\delta^{'}=1/2$. On the other hand, since the
destination knows the value of $x_1(i)$, it can cancel it by
adding $x_1(i)$ to the received signal. This converts the original
channel to an equivalent BSC with $\epsilon^{'}=0$. Hence, through
our noisy feedback approach, we obtain an equivalent wiretap BSC
with parameters $\epsilon^{'}=0,\delta^{'}=1/2$ resulting in
$$C_s^{f}=H(\delta^{'})-H(\epsilon^{'})=1.$$

\item $0<\delta<\epsilon<1/2$, $N_1$ and $N_2$ are independent.

Since $\delta<\epsilon$, we have $$C_s=0.$$ Also, $N_1$ and $N_2$
are independent, so $P_{YZ|X}=P_{Y|X}P_{Z|X}$. Then
from~\eqref{eq:cwithfin}, one can easily obtain
that~\cite{Maurer:TIT:93}
$$C_s^{p}=H(\epsilon+\delta-2\epsilon\delta)-H(\epsilon).$$

Our feedback scheme, on the other hand, achieves
$$C_s^{f}=1-H(\epsilon).$$
Since $H(\epsilon+\delta-2\epsilon\delta)\leq 1$, we have
$C_s^{f}\geq C_s^{p}$ with equality if and only if
$\epsilon+\delta-2\epsilon\delta=1/2$.

\item $0<\delta<\epsilon<1/2$ and $N_1(i)=N_2(i)+N^{'}(i)$, where
$\text{Pr}\{n^{'}(i)=1\}=(\epsilon-\delta)/(1-2\delta)$.

The main channel is a degraded version of the source-wiretapper
channel, $X\rightarrow Z\rightarrow Y$, as shown in
Figure~\ref{fig:maindegrad}.

\begin{figure}[thb]
\centering
\includegraphics[width=0.5\textwidth]{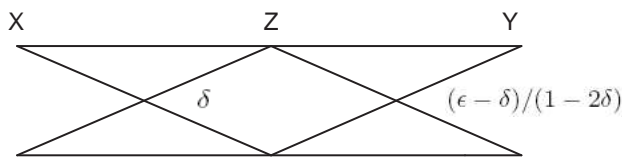}
\caption{The BSC Wiretap Channel with a Degraded Main Channel.}
\label{fig:maindegrad}
\end{figure}

Hence, from~\eqref{eq:maindegrad}, we have
\begin{eqnarray}
C_s=C_s^{p}=0,\no
\end{eqnarray}
while $C_s^{f}=1-H(\epsilon).$

 \item $0<\epsilon<\delta<1/2$, and
$N_2(i)=N_1(i)+N^{'}(i)$, where
$\text{Pr}\{n^{'}(i)=1\}=(\delta-\epsilon)/(1-2\epsilon)$.

\begin{figure}[thb]
\centering
\includegraphics[width=0.5\textwidth]{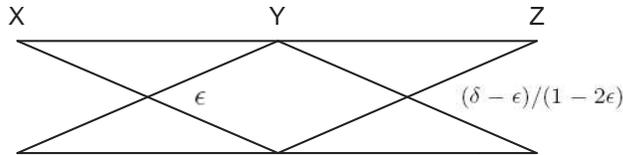}
\caption{The BSC wiretap Channel with a Degraded Source-Wiretapper
Channel.} \label{fig:wiretapdegrad}
\end{figure}

In this case, the source-wiretapper channel is a degraded version
of the main channel as shown in Figure~\ref{fig:wiretapdegrad};
$X\rightarrow Y\rightarrow Z$, so from~\eqref{eq:cwithfdeg}
\begin{eqnarray}
C_s=C_s^{p}=H(\delta)-H(\epsilon).\no
\end{eqnarray}
But $$C_s^{f}=1-H(\epsilon)\geq C_s^{p}$$ with equality if and
only if $\delta=1/2$.

\item $N_1$ and $N_2$ are correlated and the channel is not
degraded.

In this case
\begin{eqnarray}
C_s=[H(\delta)-H(\epsilon)]^+.\no
\end{eqnarray}
The value of $C_s^p$ is unknown in this case but can be bounded by
\begin{eqnarray}
C_s=[H(\delta)-H(\epsilon)]^+\leq C^p_s\leq
1-H(\epsilon)=C_s^{f}.\no
\end{eqnarray}

\end{enumerate}
In summary, the secrecy capacity with noisy feedback is always
larger than or equal to that of the public discussion paradigm
when the underlying wiretap channel is a BSC. More strongly, the
gain offered by the noisy feedback approach, over the public
discussion paradigm, is rather significant in many relevant
special cases.

\section{Even Half-duplex Feedback is Sufficient}\label{sec:half}

It is reasonable to argue against the {\em practicality} of the full
duplex assumption adopted in the previous section. For example, in
the wireless setting, nodes may not be able to transmit and receive
with the same {\em degree of freedom} due to the large difference
between the power levels of the transmit and receive chains. This
motivates extending our results to the half duplex wiretap channel
where the terminals can either transmit or receive but never both at
the same time. Under this situation, if the destination wishes to
feed back at time $i$, it loses the opportunity to receive the
$i^{th}$ symbol transmitted by the source, which effectively results
in an erasure (assuming that the source is unaware of the
destination decision). The proper feedback strategy must, therefore,
strike a balance between confusing the wiretapper and degrading the
source-destination link. In order to simplify the following
presentation, we first focus on the wiretap BSC. The extension to
arbitrary modulo-additive channels is briefly outlined afterwards.

In the full-duplex case, at any time $i$, the optimal scheme is to
let the destination send $x_1(i)$, which equals $0$ or $1$ with
probability 1/2 respectively. But in the half-duplex case, if the
destination always keeps sending, it does not have a chance to
receive information from the source, and hence, the achievable
secrecy rate is zero. This problem, however, can be solved by
observing that if at time $i$, $x_1(i)=0$, the signal the
wiretapper receives, \emph{i.e.},
$$z(i)=x_1(i)+n_2(i),$$ is the same as in the case in which the
destination does not transmit. The only crucial difference in this
case is that the wiretapper does not know whether the feedback has
taken place or not, since $x_1(i)$ can be randomly generated at
the destination and kept private.

The previous discussion inspires the following feedback scheme for
the half-duplex channel. The destination first fixes a faction
$0\leq t\leq 1$ which is revealed to both the source and
wiretapper. At time $i$, the destination randomly generates
$x_1(i)=1$ with probability $t$ and $x_1(i)=0$ with probability
$1-t$. If $x_1(i)=1$, the destination sends $x_1(i)$ over the
channel, which causes an erasure at the destination and a
potential error at the wiretapper. On the other hand, when
$x_1(i)=0$, the destination does not send a feedback signal and
spends the time on receiving from the channel. The key to this
scheme is that although the source and wiretapper know $t$,
neither is aware of the exact timing of the event $x_1=1$. The
source ignores the feedback and keeps sending information. The
following result characterizes the achievable secrecy rate with
the proposed feedback scheme.

\begin{thm}
For a BSC with half-duplex nodes and parameters $\epsilon$ and
$\delta$, the scheme proposed above achieves
\begin{eqnarray}
R_s^{f}=\max\limits_{\mu,t}\left[(1-t)\big[H(\epsilon+\mu-2\mu\epsilon)-H(\epsilon)\big]-\big[H(\hat{\delta}+\mu-2\mu\hat{\delta})-H(\hat{\delta})\big]\right]^+,
\end{eqnarray}
with $\hat{\delta}=\delta+t-2\delta t$.
\end{thm}
\begin{proof}
For the main channel, if the destination spends a $t$ fraction of
its time on sending, the equivalent main channel is shown in
Figure~\ref{fig:halfmaineq} with output $\hat{y}\in\{0,\phi,1\}$,
where $\phi$ represents an erasure. The erasure probability is
$t$. In the remaining $1-t$ fraction of the time, the channel is a
BSC with parameter $\epsilon$. Hence, the transition matrix of
this equivalent channel is

\[ \left[ \begin{array}{ccc}
(1-t)(1-\epsilon) & t & (1-t)\epsilon \\
(1-t)\epsilon & t & (1-t)(1-\epsilon)\end{array}\right].\]

Meanwhile for the wiretapper, the equivalent channel is still a
BSC, but with the increased error probability
\begin{eqnarray}
\hat{\delta}=(1-t)\delta+t(1-\delta)=\delta+t-2\delta t.
\end{eqnarray}

\begin{figure}[thb]
\centering
\includegraphics[width=0.3 \textwidth]{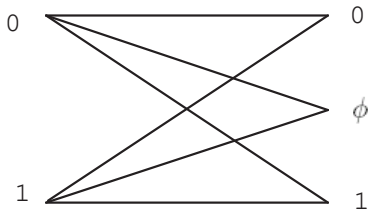}
\caption{The Equivalent Main Channel.} \label{fig:halfmaineq}
\end{figure}

Hence the original BSC wiretap channel with noisy feedback is
equivalent to a new wiretap channel $X\rightarrow (\hat{Y},Z)$
without feedback, and the channel parameters are given as above.

As shown in~\cite{Csiszar:TIT:78}, for this equivalent wiretap
channel the following secrecy rate is achievable for any input
distribution $P_X$:
\begin{eqnarray}
R^{f}=[I(X;\hat{Y})-I(X;Z)]^+.
\end{eqnarray}

Hence, by using the input distribution $\text{Pr}\{X=1\}=\mu$, one
can see that
\begin{eqnarray}
R^{f}=\max\limits_{\mu,t}\left[(1-t)\big[H(\epsilon+\mu-2\mu\epsilon)-H(\epsilon)\big]-\big[H(\hat{\delta}+\mu-2\mu\hat{\delta})-H(\hat{\delta})\big]\right]^+
\end{eqnarray}
is achievable.
\end{proof}

In general, one can obtain the optimal values of $\mu$ and $t$ by
setting the partial derivative of $R^{f}$, with respect to $\mu$
and $t$ to 0, and solving the corresponding equations.
Unfortunately, except for some special cases, we do not have a
closed form solution for these equations at the moment.
Interestingly, using the {\em not necessarily optimal} choice of
$\mu=t=1/2$, we obtain $R^{f}=(1-H(\epsilon))/2$ implying that we
can achieve a nonzero secrecy rate as long as $\epsilon\neq 1/2$
irrespective of the wiretapper channel conditions. Hence, even for
half-duplex nodes, noisy feedback from the destination allows for
transmitting information securely for {\em almost} any wiretap
BSC. Finally, we compare the performance of different schemes in
some special cases of the wiretap BSC.
\begin{enumerate}

\item $\epsilon=\delta=0$.

As mentioned above, here we have $C_s=C_s^{p}=0$. It is easy to
verify that the optimal choice of $\mu$ and $t$ are $1/2$, and we
thus have $R_s^{f}=1/2$.

\item $0<\delta<\epsilon<1/2$ and $N_1(i)=N_2(i)+N^{'}(i)$, where
$\text{Pr}\{n^{'}(i)=1\}=(\epsilon-\delta)/(1-2\delta)$.

The main channel is a degraded version of the wiretap channel, so
\begin{eqnarray}
C_s=C_s^{p}=0.
\end{eqnarray}
But by setting $\mu=t=1/2$ in our half-duplex noisy feedback
scheme, we obtain $R_s^{f}=( 1-H(\epsilon))/2.$
\end{enumerate}

The extension to the general discrete modulo-additive channel is
natural. The destination can set $\mathcal{X}_1=\mathcal{Z}$, and
generates $x_1(i)$ with certain distribution $P_{X_1}$. At time
$i$, if the randomly generated $x_1(i)\neq 0$, the destination
sends a feedback signal, incurring an erasure to itself. On the
other hand, if $x_1(i)=0$, it does not send the feedback signal
and spends the time listening to the source. The achievable
performance could be calculated based on the equivalent channels
as done in the BSC. This scheme guarantees a positive secrecy
capacity as seen in the case where $P_{X_1}$ is chosen to be
uniformly distributed over $\mathcal{Z}$. This is because a
uniform distribution over $\mathcal{Z}$ renders the output at the
wiretapper independent from the source input, i.e., $I(W;Z)=0$,
while the destination can still spend $1/|\mathcal{Z}|$ part of
the time listening to the source. Finding the optimal distribution
$P_{X_1}$, however, is tedious.
\section{The Modulo-$\Lambda$ Channel}\label{sec:exten}
In this section, we take a step towards extending our approach to
continuous valued channels. In particular, we consider the
Modulo-$\Lambda$
channel~\cite{Forney:TIT:00,Eyuboglu:TIT:92,Erez:TIT:04,ElGamal:TIT:041}.
This choice is motivated by two considerations 1) this channel
still enjoys the {\em modulo} structure which proved instrumental
in deriving our results in the discrete case, and 2) the
modulo-$\Lambda$ channel has been shown to play an important role
in achieving the capacity of the Additive White Gaussian Noise
(AWGN) channel using lattice coding/decoding
techniques~\cite{Erez:TIT:04} (in other words, an AWGN
source-destination channel can be well approximated by a
Modulo-$\Lambda$ channel). In the following, we show that, similar
to the discrete case, noisy feedback can increase the secrecy
capacity of the wiretap modulo-$\Lambda$ channel to that of the
main channel capacity in the absence of the wiretapper.

Before proceeding further, we need to introduce few more
definitions. An $m$-dimensional lattice $\Lambda\subset
\mathbb{R}^m$ is a set of points
\begin{eqnarray}
\Lambda \overset{\vartriangle}
=\{\mathbf{\lambda=\mathbf{G}\mathbf{u}:\mathbf{u}\in\mathbb{Z}}^m\},
\end{eqnarray}
where $\mathbf{G}\in \mathbb{R}^{m\times m}$ denotes the lattice
generator matrix. A fundamental region $\Omega\in \mathbb{R}^m$ of
$\Lambda$ is a set such that each $\mathbf{x}\in \mathbb{R}^m$ can
be written uniquely in the form
$\mathbf{x}=\mathbf{\lambda}+\mathbf{e}$ with
$\mathbf{\lambda}\in\Lambda,\mathbf{e}\in\Omega$, and
$\mathbb{R}^m=\Lambda+\Omega$. There are many different choices of
the fundamental region, each with the same volume which will be
denoted as $V(\Lambda)$. Given a lattice $\Lambda$, a fundamental
region $\Omega$ of $\Lambda$, and a zero-mean white Gaussian noise
process with variance $\sigma_1^2$ per dimension, the mod-$\Lambda$
channel is defined as follows~\cite{Forney:TIT:00}.

\begin{define}[\cite{Forney:TIT:00}]
The input of the mod-$\Lambda$ channel consists of points
$\mathbf{X}\in \Omega$; the output of the mod-$\Lambda$ channel is
$\mathbf{Y}=(\mathbf{X}+\mathbf{N}_1)\mod \Lambda$, where
$\mathbf{N}_1$ is an $m$-dimensional white Gaussian noise variable
with variance $\sigma_1^2$ per dimension. Hence $\mathbf{Y}$ is
the unique element of $\Omega$ that is congruent to
$\mathbf{X}+\mathbf{N}_1$.
\end{define}

In our wiretap mod-$\Lambda$ channel, the output at the wiretapper
(in the absence of feedback) is also given by
$\mathbf{Z}=(\mathbf{X}+\mathbf{N}_2)\mod \Lambda$. Here
$\mathbf{N}_2$ is an $m$-dimensional white Gaussian noise variable
with variance $\sigma_2^2$ per dimension. Similar to
Section~\ref{sec:model}, we consider noisy feedback, where the
destination sends a feedback signal $\mathbf{X}_1\in \Omega$ based
on its received signal, and the received signal at the source is
$\mathbf{Y}_{0}=(\mathbf{X}+\mathbf{X}_1+\mathbf{N}_0)\mod
\Lambda$, where $\mathbf{N}_0$ is an $m$-dimensional white
Gaussian noise with variance $\sigma_0^2$ per dimension. Now, the
received signal at the destination and wiretapper are
$\mathbf{Y}=(\mathbf{X}+\mathbf{X}_1+\mathbf{N}_1)\mod \Lambda$
and $\mathbf{Z}=(\mathbf{X}+\mathbf{X}_1+\mathbf{N}_2)\mod
\Lambda$, respectively.

For example, if $m=1$, $\Lambda=\mathbb{Z}$ is a lattice in
$\mathbb{R}$, with $[-1/2,1/2)$ being one of its fundamental
regions. With this lattice and fundamental region, the output at
the destination is then $Y=(X+X_1+N_1)\mod \Lambda=
X+X_1+N_1-\lfloor X+X_1+N_1+1/2\rfloor$, where $N_1$ is a
one-dimensional Gaussian random variable with variance
$\sigma_1^2$. Here $\lfloor x\rfloor$ denotes the largest integer
that is smaller than $x$. One can easily check that
$Y\in[1/2,1/2)$. The output at the wiretapper and source can be
written in a similar manner. This $m=1$ example can be viewed as
the continuous counterpart of the discrete channels considered in
Section~\ref{sec:full}.

Let $\mathbf{N}^{'}=\mathbf{N}_1\mod \Lambda$, and let
$f_{\Lambda,\sigma_1^2}(\mathbf{n}^{'})$ be the probability
density function of $\mathbf{N}^{'}$, one can easily verify
that~\cite{Forney:TIT:00}
\begin{eqnarray}
f_{\Lambda,\sigma_1^2}(\mathbf{n}^{'})=\sum\limits_{\mathbf{b}\in\Lambda}(2\pi\sigma_1^2)^{-\frac{m}{2}}\exp^{-||\mathbf{n}^{'}+\mathbf{b}||^2/2\sigma_1^2},\mathbf{n}^{'}\in\Omega.
\end{eqnarray}
Denote the differential entropy of the noise term $\mathbf{N}^{'}$
by $h(\Lambda,\sigma_1^2)$. Then
\begin{eqnarray}
h(\Lambda,\sigma_1^2)=-\int_{\Omega(\Lambda)}f_{\Lambda,\sigma_1^2}(\mathbf{n}^{'})\log
f_{\Lambda,\sigma_1^2}(\mathbf{n}^{'})\text{d}\mathbf{n}^{'}.
\end{eqnarray}
We are now ready to prove the following.
\begin{thm}
The secrecy capacity of mod-$\Lambda$ channel with noisy feedback
is
\begin{eqnarray}
C^{f}_s=\log
(V(\Lambda))-h(\Lambda,\sigma_1^2)\label{eq:modlamda}.
\end{eqnarray}
\end{thm}
\begin{proof}
The proof follows along the same lines as that of
Theorem~\ref{thm:dmc}. For the converse, \eqref{eq:modlamda} was
shown to be the capacity of the mod-$\Lambda$ channel with the
absence of the wiretap in~\cite{Forney:TIT:00}, which naturally
serves as an upper-bound for the secrecy capacity, as argued in
the proof of Theorem~\ref{thm:dmc}.

To achieve this secrecy capacity, the source generates length-$n$
codewords $\mathbf{x}$, with the $i$th element $\mathbf{x}(i)$
being chosen uniformly from $\Omega$. Hence each codeword
$\mathbf{x}\in\Omega^n\subset\mathbb{R}^{n\times m}$. Now, at time
$i$, the destination generates feedback signals $\mathbf{x}_1(i)$
with uniform distribution over the set $\Omega$, and thus the
feedback signal $\mathbf{X}_1$ is uniformly distributed over
$\Omega^n$. Based on the crypto lemma, for any codeword
$\mathbf{x}$ and any particular noise realization $\mathbf{n}_1$,
the length-$n$ random variable received at the wiretapper
$$\mathbf{Z}=\mathbf{x}+\mathbf{X}_1+\mathbf{n}_1\mod \Lambda,$$
is uniformly distributed over $\Omega^n$ and is independent with
$\mathbf{X}$. Hence, we have
\begin{eqnarray}
I(\mathbf{X};\mathbf{Z})=0.
\end{eqnarray}
On the other hand, with $\mathbf{X}$ uniformly distributed over
$\Omega^n$, the mutual information between ${X}$ and $\mathbf{Y}$
given $\mathbf{X}_1$ (the destination knows $\mathbf{X}_1$) is
\begin{eqnarray}
\frac{1}{n}I(\mathbf{X};\mathbf{Y}|\mathbf{X}_1)=\log
(V(\Lambda))-h(\Lambda,\sigma_1^2)\label{eq:modlamda1}.
\end{eqnarray}
So, for any $\epsilon>0$, there exists a code with rate
$R^{f}=C^{f}-\epsilon$ and $I(M;\mathbf{Z})=0$. This completes the
achievablity part.
\end{proof}

Our result for the modulo-$\Lambda$ channel sheds some light on
the more challenging scenario of the wiretap AWGN channel with
feedback. The difference between the two cases results from the
{\em modulo} restrictions imposed on the destination and
wiretapper outputs. The first constraint does not entail any loss
of generality due to the optimality of the modulo-$\Lambda$
approach in the AWGN setting~\cite{Erez:TIT:04}. Relaxing the
second constraint, however, poses a challenge because it destroys
the {\em modulo} structure necessary to hide the information from
the wiretapper (i.e., the crypto lemma needs the group structure).
In other words, if the wiretapper is not limited by the
modulo-operation then it can gain some additional information
about the source message from its observations. Therefore, finding
the secrecy capacity of the wiretap AWGN channel remains elusive
(at the moment, we can only compute achievable rates using
Gaussian noise as the feedback signal).

\section{conclusion}\label{sec:con}
In this paper, we have obtained the secrecy capacity (or achievable
rate) for several instantiations of the wiretap channel with noisy
feedback. More specifically, with a full duplex destination, it has
been shown that the secrecy capacity of modulo-additive channels is
equal to the capacity of the source-destination channel in the
absence of the wiretapper. Furthermore, the secrecy capacity is
achieved with a simple scheme in which the destination randomly
chooses its feedback signal from a certain alphabet set.
Interestingly, with a slightly modified feedback scheme, we are able
to achieve a positive secrecy rate for the half duplex channel.
Overall, our work has revealed a new encryption paradigm that
exploits the structure of the wiretap channel and uses a private key
known only to the destination. We have shown that this paradigm
significantly outperforms the public discussion approach for sharing
private keys between the source and destination.

Our results motivate several interesting directions for future
research. For example, characterizing the secrecy capacity of
arbitrary DMCs (and the AWGN channel) with feedback remains an
open problem. From an algorithmic perspective, it is also
important to understand how to exploit different channel
structures (in addition to the modulo-additive one) for encryption
purposes. Finally, extending our work to multi-user channel (e.g.,
the relay-eavesdropper channel~\cite{Lai:TIT:061}) is of definite
interest.
\bibliographystyle{ieeetr}

\begin{thebibliography}{10}

\bibitem{Shannon:BSTJ:49}
C.~E. Shannon, ``Communication theory of secrecy systems,'' {\em
Bell System
  Technical Journal}, vol.~28, pp.~656--715, Oct. 1949.

\bibitem{Wyner:BSTJ:75}
A.~D. Wyner, ``The wire-tap channel,'' {\em Bell System Technical
Journal},
  vol.~54, no.~8, pp.~1355--1387, 1975.

\bibitem{Maurer:LNCS:00}
U.~M. Maurer and S.~Wolf, ``Information-theoretic key agreement:
From weak to
  strong secrecy for free,'' {\em Lecture Notes in Computer Science},
  vol.~1807, pp.~356--373, 2000.

\bibitem{Leung:TIT:78}
S.~K. Leung-Yan-Cheong and M.~E. Hellman, ``The {G}aussian wiretap
channel,''
  {\em IEEE Trans. on Information Theory}, vol.~24, pp.~451--456, Jul. 1978.

\bibitem{Csiszar:TIT:78}
I.~Csisz$\acute{a}$r and J.~Korner, ``Broadcast channels with
confidential
  messages,'' {\em IEEE Trans. on Information Theory}, vol.~24, pp.~339--348,
  May 1978.

\bibitem{Maurer:TIT:93}
U.~M. Maurer, ``Secret key agreement by public discussion from
common
  information,'' {\em IEEE Trans. on Information Theory}, vol.~39,
  pp.~733--742, May 1993.

\bibitem{Ahlswede:TIT:93}
R.~Ahlswede and I.~Csisz$\acute{a}$r, ``Common randomness in
information theory
  and cryptography, part {I}: Secret sharing,'' {\em IEEE Trans. on Information
  Theory}, vol.~39, pp.~1121--1132, July 1993.

\bibitem{Csiszar:TIT:00}
I.~Csisz$\acute{a}$r and P.~Narayan, ``Common randomness and
secret key
  generation with a helper,'' {\em IEEE Trans. on Information Theory}, vol.~46,
  pp.~344--366, Mar. 2000.

\bibitem{Csiszar:TIT:04}
I.~Csisz$\acute{a}$r and P.~Narayan, ``Secrecy capacities for
multiple
  terminals,'' {\em IEEE Trans. on Information Theory}, vol.~50,
  pp.~3047--3061, Dec. 2004.

\bibitem{Maurer:TIT:03}
U.~M. Maurer and S.~Wolf, ``Secret key agreement over a
non-authenticated
  channel - {P}art {I}: Definitions and bounds,'' {\em IEEE Transactions on
  Information Theory}, vol.~49, pp.~822--831, Apr. 2003.

\bibitem{Maurer:TIT:031}
U.~M. Maurer and S.~Wolf, ``Secret key agreement over a
non-authenticated
  channel - {P}art {I}{I}: The simulatability condition,'' {\em IEEE
  Transactions on Information Theory}, vol.~49, pp.~832--838, Apr. 2003.

\bibitem{Maurer:TIT:032}
U.~M. Maurer and S.~Wolf, ``Secret key agreement over a
non-authenticated
  channel - {P}art {I}{I}{I}: Privacy amplification,'' {\em IEEE Transactions
  on Information Theory}, vol.~49, pp.~839--851, Apr. 2003.

\bibitem{Gopala:TIT:06}
P.~K. Gopala, L.~Lai, and H.~{El Gamal}, ``On the secrecy capacity
of fading
  channels,'' {\em IEEE Trans. on Information Theory}, Oct. 2006.
\newblock Submitted.

\bibitem{Liang:TIT:061}
Y.~Liang, H.~V. Poor, and S.~S. (Shitz), ``Secure communication
over fading
  channels,'' {\em IEEE Trans. on Information Theory}, 2006.
\newblock Submitted.

\bibitem{Bloch:TIT:06}
M.~Bloch, J.~Barros, M.~R.~D. Rodrigues, and S.~W. McLaughlin,
``Wireless
  information-theoretic security - part {I}: Theoretical aspects,'' {\em IEEE
  Trans. on Information Theory}, 2006.
\newblock Submitted.

\bibitem{Bloch:TIT:061}
M.~Bloch, J.~Barros, M.~R.~D. Rodrigues, and S.~W. McLaughlin,
``Wireless
  information-theoretic security - part {I}{I}: Practical implementation,''
  {\em IEEE Trans. on Information Theory}, 2006.
\newblock Submitted.

\bibitem{Li:ITA:07}
Z.~Li, R.~Yates, and W.~Trappe, ``Secure communication over
wireless
  channels,'' in {\em Information Theory and Application Workshop, UCSD}, Jan.
  2007.

\bibitem{Parada:ISIT:05}
P.~Parada and R.~Blahut, ``Secrecy capacity of {SIMO} and slow
fading
  channels,'' in {\em Proc. IEEE Internat. Symposium on Information Theory},
  (Adelaide, Australia), pp.~2152--2155, Sep. 2005.

\bibitem{Tekin:TIT:06}
E.~Tekin and A.~Yener, ``The {G}aussian multiple access wire-tap
channel,''
  {\em IEEE Trans. on Information Theory}, 2006.
\newblock Submitted.

\bibitem{Tekin:TIT:07}
E.~Tekin and A.~Yener, ``The {G}aussian multiple-access wire-tap
channel,''
  {\em IEEE Trans. on Information Theory}, 2007.
\newblock Submitted.

\bibitem{Liang:TIT:06}
Y.~Liang and H.~V. Poor, ``Generalized multiple access channels
with
  confidential messages,'' {\em IEEE Trans. on Information Theory}, 2006.
\newblock Submitted.

\bibitem{Liu:ISIT:06}
R.~Liu, I.~Maric, R.~D. Yates, and P.~Spasojevic, ``The discrete
memoryless
  multiple access channel with confidential messages,'' in {\em Proc. IEEE
  Internat. Symposium on Information Theory}, (Seattle, WA), July 9-14, 2006.

\bibitem{Oohama:ITW:01}
Y.~Oohama, ``Coding for relay channels with confidential
messages,'' in {\em
  Proc. IEEE Information Theory Workshop}, (Cairns, Australia), pp.~87 -- 89,
  Sept. 2-7, 2001.

\bibitem{Oohama:TIT:06}
Y.~Oohama, ``Relay channels with confidential messages,'' {\em
IEEE Trans. on
  Information Theory}, Nov. 2006.
\newblock Submitted.

\bibitem{Liu:TIT:07}
R.~Liu, I.~Maric, P.~Spasojevic, and R.~D. Yates, ``Discrete
memoryless
  interference and broadcast channels with confidential messages: Secrecy
  capacity regions,'' {\em IEEE Trans. on Information Theory}, 2007.
\newblock submitted.

\bibitem{Lai:TIT:061}
L.~Lai and H.~{El Gamal}, ``The relay-eavesdropper channel:
Cooperation for
  secrecy,'' {\em IEEE Trans. on Information Theory}, Dec 2006.
\newblock Submitted.

\bibitem{Mitrpant:TIT:06}
C.~Mitrpant, A.~Vinck, and Y.~Luo, ``An achievable region for the
{G}aussian
  wiretap channel with side information,'' {\em IEEE Trans. on Information
  Theory}, vol.~52, pp.~2181--2190, May 2006.

\bibitem{Forney:ALL:03}
G.~D. {Forney, Jr.}, ``On the role of {MMSE} estimation in
approaching the
  information-theoretic limits of linear gaussian channels: Shannon meets
  wiener,'' in {\em Proc. Allerton Conf. on Communication, Control, and
  Computing}, (Monticello, IL), 2003.

\bibitem{Forney:TIT:00}
G.~D. {Forney, Jr.}, ``Sphere-bound-achieving coset codes and
multilevel coset
  codes,'' {\em IEEE Trans. on Information Theory}, vol.~46, pp.~820--850, May
  2000.

\bibitem{Eyuboglu:TIT:92}
M.~V. Eyuboglu and G.~D. {Forney, Jr.}, ``Trellis precoding:
Combined coding,
  precoding and shaping for intersymbol interference channels,'' {\em IEEE
  Trans. on Information Theory}, vol.~38, pp.~301--314, Mar. 1992.

\bibitem{Erez:TIT:04}
U.~Erez and R.~Zamir, ``Achieving $\frac{1}{2}\log(1 +
\text{SNR})$ on the
  {AWGN} channel with lattice encoding and decoding,'' {\em IEEE Trans. on
  Information Theory}, vol.~50, pp.~2293--2314, Oct. 2004.

\bibitem{ElGamal:TIT:041}
H.~{El Gamal}, G.~Caire, and M.~O. Damen, ``Lattice coding and
decoding achieve
  the optimal diversity-vs-multiplexing tradeoff of {MIMO} channels,'' {\em
  IEEE Trans. on Information Theory}, vol.~50, pp.~968--985, June 2004.

\end{thebibliography}

\end{document}